\pdfoutput=1

\documentclass[twocolumn,twoside,nofootinbib,showpacs,prd,aps,superscriptaddress,tightenlines,preprintnumbers,10pt]{revtex4-1}

\usepackage{amsmath}
\usepackage{slashed}
\usepackage{graphicx}
\usepackage[hyperfootnotes=false,bookmarks=false]{hyperref}
\usepackage{amssymb}
\usepackage{mathrsfs}
\usepackage{bm}
\usepackage{xspace}
\usepackage{booktabs}
\usepackage{xcolor}
\usepackage{listings}
\usepackage{verbatim}
\usepackage{fancyvrb}
\usepackage{multirow}

\newcommand{\eq}[1]{Eq.~\eqref{eq:#1}}

\renewcommand{\sec}[1]{Sec.~\ref{sec:#1}}

\newcommand{\tab}[1]{Table~\ref{tab:#1}}

\newcommand{\nn}{\nonumber}

\newcommand{\df}{\mathrm{d}}
\newcommand{\img}{\mathrm{i}}

\newcommand{\De}{\Delta}

\newcommand{\cD}{\mathcal{D}}

\newcommand{\ECO}{\textsc{Eco}\xspace}
\newcommand{\FORM}{\textsc{Form}\xspace}

\allowdisplaybreaks[2]

\begin{document}

\preprint{\vbox{\hbox{NIKHEF 20-010}}}

\title{..., 83106786, 114382724, 1509048322, 2343463290, 27410087742, ... \\ Efficient Hilbert Series for Effective Theories}

\author{Coenraad B.~Marinissen}
\affiliation{Nikhef, Theory Group, Science Park 105, 1098 XG, Amsterdam, The Netherlands}
\affiliation{ITF, Utrecht University, Leuvenlaan 4, 3584 CE Utrecht, The Netherlands}

\author{Rudi Rahn}
\affiliation{Nikhef, Theory Group, Science Park 105, 1098 XG, Amsterdam, The Netherlands}
\affiliation{ITFA, University of Amsterdam, Science Park 904, 1098 XH Amsterdam, The Netherlands}

\author{Wouter J.~Waalewijn}
\affiliation{Nikhef, Theory Group, Science Park 105, 1098 XG, Amsterdam, The Netherlands}
\affiliation{ITFA, University of Amsterdam, Science Park 904, 1098 XH Amsterdam, The Netherlands}

\begin{abstract}
   We present an efficient algorithm for determining the Hilbert series of an effective theory and provide a companion code called \ECO (Efficient Counting of Operators) in \FORM. For example, the Hilbert series for the dimension 15 operators in the Standard Model Effective Theory (SMEFT) can be obtained in a minute on a single CPU core. While our implementation focusses on SMEFT, we allow for a flexible user input of the light degrees of freedom. Furthermore, gravity, as well as additional $U(1)$ global or gauge symmetries can be included.
\end{abstract}

\maketitle

%%%%%%%%%%%%%%%%%%%%%%%%%%%%%%%%%%%%%%%%%%%%%%%%%%%%%%%%%%%%%%%%%%%%%%%%%%%%%%%%
\section{Introduction}\label{sec:intro}
%%%%%%%%%%%%%%%%%%%%%%%%%%%%%%%%%%%%%%%%%%%%%%%%%%%%%%%%%%%%%%%%%%%%%%%%%%%%%%%%

An effective field theory (EFT) is appropriate to describe the effect of heavy degrees of freedom in the low energy limit, where there is insufficient energy to produce the corresponding particles on shell. These heavy degrees of freedom are not part of the EFT, and instead manifest themselves as higher-dimensional operators involving the light degrees of freedom $\phi_i$. In the absence of discovering new heavy degrees of freedom, EFTs provide a model-independent approach to search for new physics. Specific examples of EFTs that are relevant for this paper are the Standard Model EFT, both without (SMEFT) and with gravity (GRSMEFT). For an introduction to EFTs, see e.g.~Refs.~\cite{Manohar:2018aog,Cohen:2019wxr}. 

The operators in the EFT Lagrangian must be invariant under Lorentz transformations and the (gauge) symmetry group $G$, and operators related by integration by parts and equations of motion can be eliminated to obtain a minimal basis~\cite{Georgi:1991ch,Arzt:1993gz}. For SMEFT, there is a single operator at dimension 5~\cite{Weinberg:1979sa}, and the operator bases at dimension 6~\cite{Buchmuller:1985jz,Grzadkowski:2010es}, 7~\cite{Lehman:2014jma,Henning:2015alf} and 8~\cite{Lehman:2015coa,Henning:2015alf} have been constructed explicitly. 

The Hilbert series (HS) of an EFT counts the number $c_{k,r_1,...,r_N}$ of independent (higher-dimensional) operators involving $r_i$ fields $\phi_i$ and $k$ (covariant) derivatives~\cite{Lehman:2015via}
%%%
\begin{equation} \label{eq:HS}
    H(\cD,\{\phi_i\}) = \sum_{k=0}^\infty \sum_{n_1=0}^\infty ... \sum_{n_N=0}^\infty c_{k,n_1,...,n_N} \cD^k \phi_1^{n_1} ... \phi_N^{n_N}
\,.\end{equation}
%%%
Here $\phi_i$ is simply used to count the number of occurrences of the field in the operator and is not the field itself. (HS had previously been used in particle physics to count e.g.~flavor invariants~\cite{Jenkins:2009dy,Hanany:2010vu}.)

We will provide a rather efficient implementation in \FORM~\cite{Vermaseren:2000nd,Ruijl:2017dtg} for computing Hilbert Series, which we call \ECO (Efficient Counting of Operators). It follows the approach developed in Refs.~\cite{Henning:2015daa,Henning:2015alf,Henning:2017fpj}, but is orders of magnitude faster.
In table~\ref{tab:number} we list the number of operators in the Standard Model EFT of various dimensions, and the amount of time it took to calculate the HS on a single CPU core. Note that we count operators and their Hermitian conjugate (if distinct) separately.
Our \FORM implementation accompanies this paper. While our focus is on the SMEFT, in our code it is straightforward to add more light degrees of freedom, including gravity (as in Ref.~\cite{Ruhdorfer:2019qmk}), by adding a line in the \FORM input. Furthermore, additional $U(1)$ gauge symmetries (e.g.~a $Z'$) or global symmetries (e.g.~baryon number) can be added as well.

This paper is organized as follows: In \sec{hilbertseries}, we review the method for obtaining the HS of an EFT of Refs.~\cite{Henning:2015alf,Henning:2017fpj}. Restricting our discussion to EFTs in four space-time dimensions allows us to be quite explicit. We then describe the key insights underpinning our efficient implementation and a short overview of how to use the code in \sec{implementation}. We conclude in \sec{conclusions}.

\begin{table*}
  \begin{tabular}{l|ccccccccccc}
  \hline \hline
  Dimension & 5 & 6 & 7 & 8 & 9 & 10 & 11 & 12 & 13 & 14 & 15  \\ \hline
  One generation & 2 & 84 & 30 & 993 & 560 & 15456 & 11962 & 261485 & 257378 & 4614554 & 5474170 \\
  Three generations & 12 & 3045 & 1542 & 44807 & 90456 & 2092441 & 3472266 & 75577476 & 175373592 & 2795173575 & 7557369962  \\  
  Time (in seconds) & 0.01 & 0.04 & 0.07 & 0.18 & 0.40 & 1.0 & 2.1 & 5.8 & 11 & 27 & 57 \\ \hline 
  \end{tabular}
  \begin{tabular}{l|ccccc}
  \hline 
  Dimension & 16 & 17 & 18 & 19 & 20 \\ \hline
  One generation & 83106786 & 114382724 & 1509048322 & 2343463290 & 27410087742 \\ 
  Three generations & 104832630678 & 320370940524 & 3877200543051 & 13044941495798 & 141535779949640 \\  
  Time (in seconds) & 139  & 262  & 586 & 1351 & 3410 \\ \hline \hline  
  \end{tabular}
  \caption{Number of operators in the SMEFT of a given dimension with 1 or 3 generations, as well as the computing time on a single CPU core of a typical physicist's laptop.}
\label{tab:number}  
\end{table*}

%%%%%%%%%%%%%%%%%%%%%%%%%%%%%%%%%%%%%%%%%%%%%%%%%%%%%%%%%%%%%%%%%%%%%%%%%%%%%%%%
\section{Hilbert series}
\label{sec:hilbertseries}
%%%%%%%%%%%%%%%%%%%%%%%%%%%%%%%%%%%%%%%%%%%%%%%%%%%%%%%%%%%%%%%%%%%%%%%%%%%%%%%%

In this section we review the Hilbert Series technique, following the approach of  Refs.~\cite{Henning:2015daa,Henning:2015alf,Henning:2017fpj}. We start in \sec{noderivatives} by discussing the HS for operators without derivatives. We subsequently add derivatives, accounting for relations from integration by parts in \sec{ibp} and equations of motion in \sec{eom}. In \sec{gravity} we discuss how to include gravity.

%~~~~~~~~~~~~~~~~~~~~~~~~~~~~~~~~~~~~~~~~~~~~~~~~~~~~~~~~~~~~~~~~~~~~~~~~~~~~~~~
\subsection{Operators without derivatives}
\label{sec:noderivatives}
%~~~~~~~~~~~~~~~~~~~~~~~~~~~~~~~~~~~~~~~~~~~~~~~~~~~~~~~~~~~~~~~~~~~~~~~~~~~~~~~

We start by considering a single field $\phi$ transforming under the representation $R$ of the symmetry group $G$. An operator involving $n$ fields $\phi$ transforms under the (anti-)symmetric tensor product $\text{sym}^n(R)$ $(\Lambda^n(R))$ if $\phi$ is a boson (fermion), because the fields (anti-)commute. The corresponding HS can be obtained by exploiting the orthogonality of characters to project onto the trivial representation (the invariant terms)~\cite{Lehman:2015via},
%%%
\begin{align}
  H(0,u) &= \int_G\! \df \mu\, \left\{ \begin{tabular}{l l}
  $\text{PE}(\phi \chi_{R}(g))$ & $\phi$ is boson, \\
  $\text{PEF}(\phi \chi_{R}(g))$ & $\phi$ is fermion. \\
  \end{tabular}\right.
\end{align}
%%%
Here $\mu$ is the Haar measure for the group $G$. The characters for representation of $n$ copies of the field (including a $\phi^n$) are conveniently summed by the plethystic exponential (PE)~\cite{Benvenuti:2006qr,Feng:2007ur,Hanany:2014dia}
%%%
\begin{align} \label{eq:PE}
  \sum_{n=0}^\infty \phi^n \chi_{\text{sym}^{n}(R)}(g) & = \exp\biggl[ \sum_{n=1}^\infty \frac{\phi^n}{n} \chi_R(g^n) \biggr] \equiv \text{PE}\bigl(\phi \chi_{R}(g) \bigr), \nn \\
  \sum_{n=0}^\infty \phi^n \chi_{\Lambda^{n}(R)}(g) 
  &= \exp\biggl[- \sum_{n=1}^\infty \frac{(-\phi)^n}{n} \chi_R(g^n) \biggr] \nn \\
  &\equiv \text{PEF}\bigl(\phi \chi_{R}(g) \bigr)\,.
  \end{align}
%%%
For multiple fields $\phi_i$ one multiplies their plethystic exponentials before integrating over the Haar measure. Explicitly, for $N$ fields $\phi_i$ in representation $R_i$ of $G$,
%%%
\begin{align} \label{eq:HilbertWithoutDerivative}
  H(0,\{\phi_i\}) &= \int_G\! \df \mu\, \prod_{i=1}^N \left\{ \begin{tabular}{l l}
  $\text{PE}(\phi_i \chi_{R_i}(g))$ & $\phi_i$ is boson, \\
  $\text{PEF}(\phi_i \chi_{R_i}(g))$ & $\phi_i$ is fermion. \\
  \end{tabular}\right.
\end{align}
%%%

%~~~~~~~~~~~~~~~~~~~~~~~~~~~~~~~~~~~~~~~~~~~~~~~~~~~~~~~~~~~~~~~~~~~~~~~~~~~~~~~
\subsection{Integration by parts}
\label{sec:ibp}
%~~~~~~~~~~~~~~~~~~~~~~~~~~~~~~~~~~~~~~~~~~~~~~~~~~~~~~~~~~~~~~~~~~~~~~~~~~~~~~~

The discussion in \sec{noderivatives} applies to a general symmetry group $G$, including the Lorentz group. We focus on the Lorentz group in this section due to its interplay with derivatives. Exploiting the equivalence of its Lie algebra with $SU(2) \times SU(2)$, we denote the representation of $\phi$ by its spin $\ell = (\ell_1, \ell_2)$. 
To incorporate integration-by-parts (IBP) identities, we consider representations of the conformal group which act on objects like~\cite{Henning:2015daa}
%%%
\begin{equation}\label{eq:conformalRep}
    \left(\begin{array}{c}
    {\phi} \\ 
    {D_{\mu_{1}} \phi} \\ 
    {D_{\mu_{1}} D_{\mu_{2}} \phi} \\ 
    {\vdots}\end{array}\right).
\end{equation}
%%%
Here $\phi$ is the highest weight of this representation with scaling dimension
$\Delta$. The (covariant) derivatives act as a lowering operators, so IBP identities can be
taken into account by solely focussing on the highest weight state. Derivatives
transform as $(\tfrac12,\tfrac12)$ under the Lorentz group and their commutators
yield field strength (and Weyl) tensors when including gauge groups (gravity), so the character of the representation of
\eq{conformalRep} is
%%%
\begin{align} \label{eq:char}
    \chi_{\Delta,l} (q,x) 
    &= q^\Delta \chi_\ell (x) \sum_{n=0}^{\infty} q^{n} \chi_{\text{sym}^{n}(\frac{1}{2},\frac{1}{2})}(x)
    \nn \\ &
    \equiv q^\Delta \chi_l (x) P(q,x)\,.
\end{align}
%%%
Here $x$ parametrizes the Lorentz group with its Haar measure $\df \mu_{L}$,
and $q$ denotes the scaling dimension, that enter in the orthogonality of
characters\footnote{For modifications due to equations of motions see Ref.~\cite{Henning:2017fpj}.}
%%%
\begin{align} \label{eq:ortho}
 \int\! \df \mu_{L} \oint\! \frac{\df q}{2\pi \img\, q}\frac{1}{|P(q,x)|^2}\,
 \chi_{\Delta,\ell}^* \chi_{\Delta',\ell'}  &= \delta_{\ell,\ell'} \delta_{\Delta,\Delta'}.
\end{align}
%%%

To construct the HS for the Lorentz group we need the character for multiple
fields, which is  given by the plethystic exponential in \eq{PE} of the
character in \eq{char}. In this case we don't project onto the trivial
representation, but rather onto Lorentz scalars with a specific scaling
dimension. Since we want the highest weight of this representation, we can
directly read off the number of derivatives, by dividing out the scaling
dimension of the field. This leads to the following expression for the HS of a
bosonic field, which takes into account IBP identities,
%%%
\begin{align} 
  H(\cD,\phi) &= 1 + \sum_k  \cD^k \!\!
   \int\! \df \mu_{L} \oint\! \frac{\df q}{2\pi \img\, q}\frac{1}{|P(q,x)|^2}\,
   \\ & \quad \times
  \chi_{k,(0,0)}^* \Bigl[\text{PE}\Bigl(\frac{\phi}{q^\De} \chi_{\Delta,\ell}(q,x) \Bigr) -1\Bigr] \nn \\
  &= 1 + \!\! \int\! \df \mu_{L} \frac{1}{P(\cD,x)}\,
   \Bigl[\text{PE}\Bigl(\frac{\phi}{q^\De} \chi_{\Delta,\ell}(\cD,x) \Bigr) -1\Bigr]  
\,.\nn\end{align}
%%%
Here $\cD$ counts the number of (covariant) derivatives. 
To obtain the last line the integral over $q$ is performed. For a fermionic field the PE is replaced by PEF. The generalization to multiple fields again involves multiplying the corresponding plethystic exponentials. To add symmetry groups, the corresponding characters and Haar measures should be included.

%~~~~~~~~~~~~~~~~~~~~~~~~~~~~~~~~~~~~~~~~~~~~~~~~~~~~~~~~~~~~~~~~~~~~~~~~~~~~~~~
\subsection{Equations of motion}
\label{sec:eom}
%~~~~~~~~~~~~~~~~~~~~~~~~~~~~~~~~~~~~~~~~~~~~~~~~~~~~~~~~~~~~~~~~~~~~~~~~~~~~~~~

EOM redundancies can be removed by modifying the characters for the conformal
representation in the preceding section. We start with the EOM for a Lorentz
scalar, which allows us to drop $D^2 \phi$. Contracted indices
correspond to taking a trace, which we therefore need to remove from \eq{conformalRep}. For example, for the row with two derivatives in \eq{conformalRep}, 
%%%
\begin{equation}\label{eq:EOMexplicit}
D_{\mu_1} D_{\mu_2} \phi = (D_{\mu_1} D_{\mu_2} -  \tfrac14 \eta_{\mu_1\mu_2} D^2) \phi^2 +  \tfrac14 \eta_{\mu_1\mu_2} D^2 \phi^2
\,,\end{equation}
%%%
we only want the first term on the right-hand side. 
While the generalization of \eq{EOMexplicit} to the row with $n$ derivatives is cumbersome, the corresponding character under the Lorentz group is 
%%%
\begin{equation}\label{eq:EOMscalarchar}
\chi_{\text{symm}^n(\frac{1}{2},\frac{1}{2})}(x) = \tilde{\chi}(x) +
\chi_{\text{symm}^{n-2}(\frac{1}{2},\frac{1}{2})}(x) \cdot 1,
\end{equation}
%%%
with 1 the trivial character of $D^2 \phi$ and $\tilde{\chi}$ the character of the traceless part that we seek. Note that this also removes all powers of equations of motion, since these are descendants of $D^2 \phi$.
Implementing this modified Lorentz character changes the character for the conformal
representation to
%%%
\begin{align} \label{eq:EOMconfscal}
    \tilde{\chi}_{\Delta,l} (q,x) 
    &= q^\Delta (1-q^2) \chi_l (x) P(q,x)\,.
\end{align}
%%%

For fermions and gauge fields the Lorentz structure of the fields enters the
fray. The EOM for a lefthanded Weyl fermion allows us to remove $\slashed{D}\psi$,
which transforms as $(0,1/2)$ under the Lorentz group, corresponding to the first term on the right-hand side in the decomposition of the tensor product
%%%
\begin{equation}\label{eq:EOMfermiondecomp}
\Bigl(\frac{1}{2},\frac{1}{2}\Bigr)\otimes\Bigl(\frac{1}{2},0\Bigr) =
\Bigl(0,\frac{1}{2}\Bigr)\oplus \Bigl(1,\frac{1}{2}\Bigr)
\,.\end{equation}
%%%
Accordingly, the Lorentz character corresponding to $n$ derivatives acting on a
left-handed spinor decomposes as
%%%
\begin{equation}\label{eq:EOMcharfermion}
\chi_{(\frac{1}{2},0)} \chi_{\text{symm}^n(\frac{1}{2},\frac{1}{2})}
= \tilde{\chi} +
\chi_{\text{symm}^{n-1}(\frac{1}{2},\frac{1}{2})}\chi_{(0,\frac{1}{2})},
\end{equation}
%%%
suppressing the argument $x$ of these characters.
Using $\tilde\chi$ instead of $\chi$ modifies the character for the conformal
representation to
%%% 
\begin{equation}\label{eq:EOMconffermion}
\tilde{\chi}_{\Delta,l} (q,x)=q^\Delta \bigl(\chi_{(\frac{1}{2},0)}(x) - q
\chi_{(0,\frac{1}{2})}(x)\bigr) P(q,x),
\end{equation}
%%%
and similarly for right-handed fermions.

Gauge fields transform under $(1,0)\oplus(0,1)$, hence it is easiest to look at
the self-dual and anti-self-dual, $(1,0)$ and $(0,1)$, first. We note here that 
due to the identity $D^2 F^{\mu\nu}=0$, which follows from the Bianchi identity, we only consider 
symmetrized and traceless combinations of derivatives\footnote{See the discussion on \emph{short representations} in~\cite{Henning:2017fpj}.}.
A derivative acting
on a $(1,0)$ then decomposes as
%%%
\begin{equation}\label{eq:EOMgaugedecomp}
\Bigl(\frac{1}{2},\frac{1}{2}\Bigr)\otimes\bigl(1,0\bigr) =
\Bigl(\frac{1}{2},\frac{1}{2}\Bigr)\oplus \Bigl(\frac{3}{2},\frac{1}{2}\Bigr),
\end{equation}
%%%
and we identify the first term on the RHS with the EOM, $D_\mu F^{\mu\nu}=J^\nu$. Since currents are conserved, we'd overreach by
removing the full vector component, and have to separate out the trace $D_\mu J^\mu = 0$.
The character for the conformal representation is therefore given by
%%%
\begin{equation}\label{eq:EOMconfgaugepart}
\tilde{\chi}_{\Delta,l} =q^\Delta \bigl(\chi_{(1,0)} - q
\chi_{(\frac{1}{2},\frac{1}{2})} + q^2\bigr) P(q,x),
\end{equation}
%%%
where we again suppressed the arguments of the characters. A similar argument holds
for the $(0,1)$ component of the field strength. Combining these, we find that the character of the conformal representation for the field strength after removing EOM is given by
%%%
\begin{equation}\label{eq:EOMconfgauge}
\tilde{\chi}_{\Delta,l}=q^\Delta \Big(\chi_{(1,0)\oplus(0,1)} - 2q
\chi_{(\frac{1}{2},\frac{1}{2})} + 2q^2\Big) P(q,x).
\end{equation}
%%%

%~~~~~~~~~~~~~~~~~~~~~~~~~~~~~~~~~~~~~~~~~~~~~~~~~~~~~~~~~~~~~~~~~~~~~~~~~~~~~~~
\subsection{Gravity}
\label{sec:gravity}
%~~~~~~~~~~~~~~~~~~~~~~~~~~~~~~~~~~~~~~~~~~~~~~~~~~~~~~~~~~~~~~~~~~~~~~~~~~~~~~~

We will now discuss how to include gravity, providing a brief summary of the approach in Ref.~\cite{Ruhdorfer:2019qmk}.
 Quantizing the action of general relativity (GR) 
%%%
\begin{equation}\label{eq:einsteinHilbert}
S = -\frac{1}{16\pi G}\int\! \df^4x\, \sqrt{-g} R
\end{equation}
%%%
yields a non-renormalizable theory. This can of course be used as an
effective theory, for which we must include operators of higher mass dimension
in the Riemann tensor $R_{\mu\nu\rho\sigma}$, whose reducible
representation under the Lorentz group is
$(2,0)\oplus(0,2)\oplus(1,1)\oplus(0,0)$. To see which EOM
redundancies we need to remove, we look at the Einstein equations
%%%
\begin{equation}\label{eq:einsteinEquations}
R_{\mu\nu} - \frac{1}{2}g_{\mu\nu} R = 8\pi G T_{\mu\nu},
\end{equation}
%%%
with $R_{\mu\nu}$ and $R$ the Ricci tensor and scalar, respectively. In vacuum
the energy momentum tensor $T_{\mu\nu}$ is zero and the Einstein equations
reduce to $R_{\mu\nu}=0$. As explained in Ref.~\cite{Ruhdorfer:2019qmk}, we can
perform field redefinitions in the metric tensor to remove any occurrence of
Ricci scalar and tensor from the operator basis. Therefore, the building block
in the EFT of gravity is the Weyl tensor $C_{\mu\nu\rho\sigma}$, which is the
Riemann tensor sans its Ricci tensor/scalar traces. The representation of
$R_{\mu\nu}$ under the Lorentz group is $(1,1)\oplus(0,0)$, meaning that the
Weyl tensor has to transform as $(2,0)\oplus(0,2)$. Besides satisfying the Einstein
equations, we have to take the contracted Bianchi identity
%%%
\begin{equation}\label{eq:bianchi}
\nabla^\mu C_{\mu\nu\rho\sigma} = \nabla_{[\rho}R_{\sigma]\nu} + \frac{1}{6}g_{\nu[\rho}\nabla_{\sigma]}R
\end{equation}
%%%
into account, which can be regarded as an additonal equation of motion. 

To see
which expressions we must subtract in the character for $C_{\mu\nu\rho\sigma}$,
we again work with the self-dual and anti-self-dual of the Weyl tensor, which yields the irreducible subspaces $(2,0)$ and $(0,2)$.
As in the gauge field case, commutators of derivatives yield building blocks
that are already included, and the Bianchi identity implies that $\nabla^2
C_{\mu\nu\rho\sigma}$ is not an independent quantity, so we again consider
symmetrized and traceless products of derivatives. The tensor product of a derivative acting on $(2,0)$ decomposes as
%%%
\begin{equation}\label{eq:bianchiTensorProduct}
\Bigl(\frac{1}{2},\frac{1}{2}\Bigr) \otimes \Bigl(2,0\Bigr) = \Bigl(\frac{5}{2},\frac{1}{2}\Bigr)\oplus\Bigl(\frac{3}{2},\frac{1}{2}\Bigr)
\end{equation}
%%%
and we identify $(\frac{3}{2},\frac{1}{2})$ with the contracted Bianchi
identity. Similar to the gauge field case, subtracting the full
$(\frac{3}{2},\frac{1}{2})$ removes too much, as $\nabla^\mu\nabla^\nu
C_{\mu\nu\rho\sigma}\equiv0$ by antisymmetry. This vanishing object is an
antisymmetric rank-2 tensor, and so transforms as $(1,0)$. Therefore, the
character of the conformal representation is given by
%%%
\begin{equation}\label{eq:EOMconfWeylpart}
\tilde{\chi}_{\Delta,l} =q^\Delta \bigl(\chi_{(2,0)} - q
\chi_{(\frac{3}{2},\frac{1}{2})} + q^2\chi_{(1,0)}\bigr) P(q,x),
\end{equation}
%%%
and we find a similar expression for the $(0,2)$ component. Combining these
results yields the EOM-removed character for the conformal
representation of the Weyl tensor as
%%%
\begin{align}\label{eq:EOMconfWeyl}
\tilde{\chi}_{\Delta,l} &=q^\Delta \, P(q,x) \, \Big(\chi_{(2,0)\oplus(0,2)}
\nonumber \\
&\quad - q \chi_{(\frac{3}{2},\frac{1}{2})\oplus(\frac{1}{2},\frac{3}{2})} +
2q^2\chi_{(1,0)\oplus(0,1)}\Big) .
\end{align}
%%%

%%%%%%%%%%%%%%%%%%%%%%%%%%%%%%%%%%%%%%%%%%%%%%%%%%%%%%%%%%%%%%%%%%%%%%%%%%%%%%%%
\section{Implementation}
\label{sec:implementation}
%%%%%%%%%%%%%%%%%%%%%%%%%%%%%%%%%%%%%%%%%%%%%%%%%%%%%%%%%%%%%%%%%%%%%%%%%%%%%%%%
In this section we discuss \ECO, our implementation of the Hilbert series in \FORM. We start in \sec{algorithm} by discussing the structure of the algorithm, explaining some of methods that are key to making it an efficient implementation. In \sec{code} we provide instructions on how the code can be used, and how it can be applied to different EFTs. The possibility to add local and/or global $U(1)$ symmetries is discussed in \sec{U1}. In \sec{results} we illustrate the use of our program by applying it to SMEFT, SMEFT for a Higgs-doublet model and GRSMEFT, reproducing known results and obtaining new ones at higher dimensions.

%~~~~~~~~~~~~~~~~~~~~~~~~~~~~~~~~~~~~~~~~~~~~~~~~~~~~~~~~~~~~~~~~~~~~~~~~~~~~~~~
\subsection{Structure of the algorithm}
\label{sec:algorithm}
%~~~~~~~~~~~~~~~~~~~~~~~~~~~~~~~~~~~~~~~~~~~~~~~~~~~~~~~~~~~~~~~~~~~~~~~~~~~~~~~
Using the Hilbert series in \sec{hilbertseries}, we see that implementing an algorithm to enumerate the operators in a minimal basis at a desired mass dimension is straightforward. It amounts to inserting the explicit form of the characters for all the different fields, expanding the plethystic exponentials at that mass dimension and integrating over the Lorentz group and the gauge groups. In principle, integrating can be a hard problem, but from the form of the characters and the Haar measure in \tab{characters} we see that this amounts to the simple task of taking residues. However, if this is implemented without any refinement, it results in an inefficient algorithm that generates a huge number of terms, of which many will be zero in the end. The trick to getting an efficient implementation is keeping the number of terms in the expansion as small as possible, and figuring out terms that will be zero before carrying out the whole expansion.

To illustrate how to keep the number of terms small, we discuss the example of the Hilbert series for multiple left-handed fermions $\psi_i$ ($\De = \frac32$ and $\ell=(\frac12,0)$), which are all charged differently under the gauge group $G$
%%%
\begin{align}
&H(\cD,\phi) 
\\
&=\! 1 \!+ \!\! \int\! \df \mu_{L,G} \frac{1}{P(\cD,x)}\,
\Bigl[\prod_{i=1}\text{PEF}\Bigl(\frac{\psi_i}{q^\De} \tilde{\chi}_{\De,\ell}(\cD,x)\chi_i \Bigr) -1\Bigr] \nn \\
&=\! 1 \!+ \!\! \int\! \df \mu_{L,G} \frac{1}{P(\cD,x)} \Bigl[\text{PEF}\Bigl(q^{-\De}\tilde{\chi}_{\De,\ell}(\cD,x) \! \sum_i \!\psi_i\chi_i  \Bigr) \!-\!1\Bigr]
,\nn\end{align}
%%%
with $\chi_i$ the character of the representation of $\psi_i$ under $G$. Because the only dependence on the mass dimension in the argument of the plethystic exponential is in $q^{-\De}\chi_{\De,\ell}(\cD,x)$, we see that $\sum_i\psi_i\chi_i$ can be treated as one element during the expansion in mass dimension, and the only important feature is the power of its argument (this arises from the $g^n$ in \eq{PE}). 

To efficiently expand the plethystic exponential, which involves many terms, we make great use of the \texttt{Brackets+} and \texttt{id, once} features of \FORM. With the \texttt{Brackets+} statement we can order the expression with all terms of equal mass dimension in a single bracket. The \texttt{id, once} statement tries to match terms one by one with another expression, giving \FORM the possibility to sort the whole expression after each insertion i.e. combining same terms. The interested reader can have a look at the comments in the code to see how we used these \FORM statements for an efficient expansion. Furthermore, we can postpone inserting the explicit form of $\chi_{(\frac{1}{2},0)}$ and $\chi_{(0,\frac{1}{2})}$ until after the expansion in mass dimension (once again, you only have to keep track of the power of the variables of these characters), when we integrate over the Lorentz group, thus reducing the number of terms in intermediate expressions.

From the character, which is the product of the Lorentz character with the character of the gauge group, we note that after the expansion we can first perform the integral over the Lorentz group and still treat $\sum_i\psi_i\chi_i$ just as one term. The integral over the Lorentz group will set many terms equal to zero, so it is not until we perform the integral over the gauge group, that we have to expand the sum $\sum_i\psi_i\chi_i$. Of course, when the gauge group is the product of other groups, the character is a product of characters meaning we can use this trick again; only inserting explicit expressions for these characters when we perform the integral over the corresponding group. Inserting the characters of the gauge groups can be done efficiently by making use of \texttt{id, once} again.  

Therefore, the algorithm has to perform one expansion for the plethystic exponential of a scalar field, fermion, Field strength or gravity tensor and for every field type a commuting function\footnote{If all objects in a power are commuting, \FORM makes use of binomial expansions, making the expansion a lot faster.} is used that can be replaced later by the field content (i.e. the symbol to count the occurrence of a field and its character under the gauge group). Furthermore, to account for IBP relations, we have to expand the prefactor $\frac{1}{P(\cD,x)}$. From \eq{char}, we see that this is a plethystic exponential itself,
%%%
\begin{equation}
\frac{1}{P(\cD,x)} = \text{PE}[-\cD \chi_{(\frac{1}{2},\frac{1}{2})}(x)],
\end{equation}
%%%
and in order to expand this we can follow the recipe of subsequent substitutions, as described above.

For the Standard Model gauge group $U(1)\times SU(2)\times SU(3)$, our implementation takes on the following form:
\begin{itemize}
\item Read in which fields are present, and store their representation under the symmetry groups.
\item Expand just one plethystic exponential for every type of field (scalar, fermion, field strength or gravity), and multiply these to get the Hilbert series.
\item Insert expressions for Lorentz characters, and perform integral over Lorentz group (from residues).
\item Insert expressions for gauge group characters and perform integral, one at the time. First for $SU(2)$, then $U(1)$, and finally $SU(3)$.
\end{itemize}
 
Another trick we employ is that we can express characters in terms of a basis. For example, $2\otimes 2 = 3 \oplus 1$, so $\chi_3 = \chi_2^2 - 1$. This results in a faster algorithm as we can now use the power of \FORM to combine terms, and only need to substitute the characters of our basis. For $SU(2)$, and similarly the Lorentz group, we therefore only require the character for spin $\frac{1}{2}$.

%~~~~~~~~~~~~~~~~~~~~~~~~~~~~~~~~~~~~~~~~~~~~~~~~~~~~~~~~~~~~~~~~~~~~~~~~~~~~~~~
\subsection{How to use ECO}
\label{sec:code}
%~~~~~~~~~~~~~~~~~~~~~~~~~~~~~~~~~~~~~~~~~~~~~~~~~~~~~~~~~~~~~~~~~~~~~~~~~~~~~~~

\begin{table*}
	\begin{tabular}{l|p{10cm}}
		\hline \hline
		Procedure & Description \\
		\hline \hline
		\texttt{add`Field'(`symbol',`SU3',`SU2',`Q')} & Adds the field with its \texttt{symbol} (\texttt{symbol} needs to be declared first in \FORM) to the Hilbert series. For an efficient algorithm to only count the operators replace symbol by 1, that is use \texttt{\#call add`Field'(1,`SU3',`SU2',`Q')}. See \tab{characters} for an overview of the different fields that can be used and what form is expected for the input of the symmetry group $SU(3)\times SU(2)\times U(1) $. \\
		\hline
		\texttt{HilbertSeries(`symbol')} & Computes the HS at mass dimension \texttt{massDim} with \texttt{symbol} the symbol for the derivative (\texttt{symbol} needs to be declared first in \FORM). Needs to be called after all particles are added with \texttt{add`Field'}\\
		\hline
		\texttt{counting} & Counts the number of operators in the output of \texttt{HilbertSeries}. Needs to be called after \texttt{HilbertSeries}.\\
		\hline  \hline
	\end{tabular}
	\caption{Overview of all procedures}
	\label{tab:procedures}
\end{table*}

The user can specify the input in the file \texttt{main.frm}, and execute \ECO by calling \texttt{form main} or \texttt{tform -wn main}, where $n$ denotes the number of cores. The structure of this file is as follows: the first part contains settings, such as the desired mass dimension, after which the different fields of the model are specified. The final part of the file performs the calculation described above, giving the Hilbert Series as the output. A summary of all procedures and their action can be found in \tab{procedures}. All the declarations the program needs and the procedures that can be used to add fields are in the files \texttt{declare.h}, \texttt{addField.h}, and \texttt{HilbertSeries.h}, respectively. These files are included as header files in the \texttt{main} file.

We start by discussing the settings, of which the most important one is the desired mass dimension, which one specifies using the variable \texttt{massDim}. One can choose whether or not EOM and IBP relations should be used to reduce the basis by setting \texttt{EOM} and \texttt{IBP} to \texttt{1} or \texttt{0}, respectively. As an additional feature (useful e.g.~in SMEFT),  the number of Fermion generations can be defined with \texttt{numFermGen} (by default this is 1). For example, when we want to generate a basis at mass dimension 6, subtract both EOM and IBP relations and with one fermion generation, the \texttt{Settings} section in the main file includes

\begin{Verbatim}[commandchars=\\\{\}]
 #define massDim "6"
 #define EOM "1"
 #define IBP "1"
 #define numFermGen "1"
  \vdots
\end{Verbatim}

Next we specify the fields by calling the procedure 
%%%
\begin{verbatim}
 #call add`Field'(`symbol',`SU3',`SU2',`Q')
\end{verbatim}
%%%
for every field separately. Here \texttt{`Field'} refers to the transformation under the Lorentz group, e.g.~scalars or left-handed fermions, and a complete list of all supported particles is given in \tab{characters}. The argument \texttt{Symbol} of this procedure encodes the symbol used to denote the field, which needs to be declared before calling the procedure.\footnote{All protected symbols can be found in the \texttt{declare.h} file, the most important of which are \texttt{x,y,y1,y2,z1,z2}.} The next three parameters are the representations under $SU(3) \times SU(2) \times U(1)$. For $SU(3)$ and $SU(2)$ the input of the representation is equal to the dimension of the representation. For the representation under SU(3) the possibilities are the singlet (\texttt{1}), $3$ (\texttt{3}), $\bar{3}$ (\texttt{3B}) and $8$ (\texttt{8}) representations, and for $SU(2)$ one can choose between the singlet (\texttt{1}), doublet (\texttt{2}) and triplet (\texttt{3}) representations. The charge under $U(1)$ needs to be an integer \texttt{Q}, which we achieve by rescaling with a factor of 6 (similarly we rescaled the mass dimensions in our internal code such that they are integers). Charges for additional $U(1)$ symmetries can be added at the end of the string (not shown), as discussed in \sec{U1}.

An example in which the Higgs field and the left-handed quark doublet of the SM are added looks as follows
%%%
\begin{Verbatim}[commandchars=\\\{\}]
  \vdots
 Symbol h,hd,Q,Qd;

 #call addScalar(h,1,2,3)
 #call addScalar(hd,1,2,-3)

 #call addLHFermion(Q,3,2,1)
 #call addRHFermion(Qd,3B,2,-1)
  \vdots
\end{Verbatim}
%%%
Note that the conjugate particles need to be added as independent building blocks. 

When all fields are declared, the HS is computed by calling the \texttt{HilbertSeries} procedure, which takes the symbol used for momentum as its argument. To count operators one can set the argument to 1, and when set to 0 only operators without derivatives are produced. This procedure carries out the calculation discussed before and therefore takes up the bulk of the run time. The output of this procedure is a \texttt{Local} expression \texttt{Hilbert} that gives the basis as a polynomial of the symbols of the fields declared above and the user-specified symbol for the derivative. For the above example this yields
\begin{Verbatim}[commandchars=\\\{\}]
 \vdots
 Symbol p;
 #call HilbertSeries(p)
 Print;
 .sort

\textcolor{blue}{Hilbert = 2*Q^2*Qd^2 + 2*h*hd*Q*Qd*p} 
 \textcolor{blue}{+ 2*h^2*hd^2*p^2 + h^3*hd^3;}

 #call counting
\textcolor{blue}{Number of operators at mass dimension 6 is 7.}

 .end
\textcolor{blue}{0.03 sec out of 0.03 sec}
\end{Verbatim}
%%%
where we also display the result of the \FORM program in blue. The number of operators can be counted and printed by calling the \texttt{counting} procedure. This procedure only prints the total number of operators that are in the expression \texttt{Hilbert}, without overwriting it.

%~~~~~~~~~~~~~~~~~~~~~~~~~~~~~~~~~~~~~~~~~~~~~~~~~~~~~~~~~~~~~~~~~~~~~~~~~~~~~~~
\subsection{Additional U(1) symmetries and generations}
\label{sec:U1}
%~~~~~~~~~~~~~~~~~~~~~~~~~~~~~~~~~~~~~~~~~~~~~~~~~~~~~~~~~~~~~~~~~~~~~~~~~~~~~~~
In addition to the gauge groups of the SM, we have included the possibility to add one (or more)  $U(1)$ symmetry group(s). Such a $U(1)$ can be either an extra gauge symmetry, e.g.~a $Z'$ model, or a global symmetry such as baryon number. From the point of view of enumerating operators, the only difference between a gauge and global symmetry is that in the former case one needs to include the corresponding field strength in the list of particles. The charges under these additional $U(1)$ symmetries can be added as extra arguments when listing particles. E.g.~for baryon number,
%%%
\begin{Verbatim}[commandchars=\\\{\}]
  \vdots
 #call addLHFermion(Q,3,2,1,1)
 #call addRHFermion(Qd,2B,2,-1,-1)
  \vdots
\end{Verbatim}
%%%
If for a field no charge corresponding to an additional $U(1)$ is provided, we assume that it has charge 0. We remind the reader that these charges must be integers, which is why we multiplied baryon number by a factor of 3. 

To get general dependence on the number of fermion generations, we can declare a symbol for this (at the beginning of the \texttt{main} file)
%%%
\begin{Verbatim}[commandchars=\\\{\}]
 Symbol Nf;
 #define numFermGen "Nf"
  \vdots
 #call HilbertSeries(p)
 Print;
 .sort

\textcolor{blue}{Hilbert = 2*h^2*hd^2*p^2 + h^3*hd^3} 
 \textcolor{blue}{+ Nf^2*Q^2*Qd^2 + 2*Nf^2*h*hd*Q*Qd*p}
 \textcolor{blue}{+ Nf^4*Q^2*Qd^2;}
\end{Verbatim}
%%%
To organize this expression in powers of \texttt{Nf} one can use the \texttt{Brackets} command. Details on its use and other tips can be found in the \ECO package.

%~~~~~~~~~~~~~~~~~~~~~~~~~~~~~~~~~~~~~~~~~~~~~~~~~~~~~~~~~~~~~~~~~~~~~~~~~~~~~~~
\subsection{Results}
\label{sec:results}
%~~~~~~~~~~~~~~~~~~~~~~~~~~~~~~~~~~~~~~~~~~~~~~~~~~~~~~~~~~~~~~~~~~~~~~~~~~~~~~~

With \ECO it is now straightforward to reproduce known results for (extensions of) SMEFT. The results for SMEFT up to mass dimension 15 were already given in \cite{Henning:2015alf} and with the \FORM code we extended this up to dimension 20 in a reasonably short amount of time, see \tab{number}. We reproduced the results of Ref.~\cite{Ruhdorfer:2019qmk} for GRSMEFT at dimension 8 and extended it to dimension 15, see \tab{numberGR}. As an illustration of including an additional $U(1)$ symmetry, we counted how many of the operators in SMEFT conserve baryon minus lepton number ($B-L$). None of the operators of odd dimension do, while all of the operators of even dimension do up to and including dimension 8 (10) for three (one) generations of fermions. An example of an operator of dimension 10 that violates $B-L$ for two or more generations is $h^4 \ell_1^2 \ell_2^2$. This operator must vanish for $\ell_1 = \ell_2$ due to the antisymmetry of fermion fields. Our full results are listed in \tab{numberBL}. Calculating the HS for GRSMEFT is about a factor two slower than SMEFT, while obtaining the $B-L$ conserving operators is about 10\% faster.

The Hilbert series approach was applied to the Two Higgs Doublet Model (2HDM) in Ref.~\cite{Anisha:2019nzx}. In this extension of the SM, an identical second Higgs doublet is added. We reproduced the 228 operators found in \cite{Anisha:2019nzx} at mass dimension 6. As most of the operators in the SM have a coupling to the Higgs field, it is not surprising that we find a sizable number of additional operators for the 2HDM. This has of course implications for the run time of the program. At mass dimension 15 we find a number of 22020182 (16181746764) operators with one (three) generation(s) respectively. Producing the full operator basis results in a run time which is a factor 2 slower compared to SMEFT. For counting the total number of operators only, more terms can of course be combined, giving a run time which is just a few percent slower. 

\begin{table*}
\begin{tabular}{l|ccccccccccc}
	\hline \hline
	Dimension & 5 & 6 & 7 & 8 & 9 & 10 & 11 & 12 & 13 & 14 & 15  \\ \hline
	One generation & 2 & 94 & 30 & 1096 & 580 & 17797 & 12936 & 314650 & 291702 & 5812440 & 6518462 \\
	Three generations & 12 & 3055 & 1542 & 45816 & 91284 & 2160964 & 3567228 & 79514441 & 182542620 & 2995340275 & 8023911776  \\ \hline \hline
\end{tabular}
\caption{Number of operators in the GRSMEFT of a given dimension with 1 or 3 generations.}
\label{tab:numberGR}  
\end{table*}

\begin{table*}
	\begin{tabular}{l|ccccccccccc}
		\hline \hline
		Dimension & 5 & 6 & 7 & 8 & 9 & 10 & 11 & 12 & 13 & 14 & 15  \\ \hline
		One generation & 0 & 84 & 0 & 993 & 0 & 15456 & 0 & 261421 & 0 & 4612082 & 0 \\
		Three generations & 0 & 3045 & 0 & 44807 & 0 & 2091965 & 0 & 75497816 & 0 & 2788483269  & 0 \\ \hline \hline
	\end{tabular}
	\caption{Number of operators in the SMEFT that conserve baryon minus lepton number.}
	\label{tab:numberBL}  
\end{table*}

%%%%%%%%%%%%%%%%%%%%%%%%%%%%%%%%%%%%%%%%%%%%%%%%%%%%%%%%%%%%%%%%%%%%%%%%%%%%%%%%
\section{Conclusions}
\label{sec:conclusions}
We have presented an efficient algorithmic implementation of the Hilbert series
for the SMEFT with some extensions, and release a \FORM instantiation of this
approach called \ECO. The speed-up due to how we structured the calculation
 and from using \FORM allows us to determine the number of operator and field content
of a minimal basis up to mass dimensions as high as 20, in less than an hour on a single core.
We supplemented our code with capabilities for spin-2 fields and additional
$U(1)$ symmetries, which extends its application to GRSMEFT and models with
additional $Z'$ bosons, or global symmetries such as baryon or lepton number,
among others. As we also allow for the inclusion of additional light scalars,
vectors, and fermions, we expect this tool to be a valuable addition to the
model-building toolkit.
%%%%%%%%%%%%%%%%%%%%%%%%%%%%%%%%%%%%%%%%%%%%%%%%%%%%%%%%%%%%%%%%%%%%%%%%%%%%%%%%

%~~~~~~~~~~~~~~~~~~~~~~~~~~~~~~~~~~~~~~~~~~~~~~~~~~~~~~~~~~~~~~~~~~~~~~~~~~~~~~~
\acknowledgements
%~~~~~~~~~~~~~~~~~~~~~~~~~~~~~~~~~~~~~~~~~~~~~~~~~~~~~~~~~~~~~~~~~~~~~~~~~~~~~~~

We thank Jos Vermaseren for assistance with \FORM. This work is supported by the ERC grant ERC-STG-2015-677323, the NWO projectruimte 680-91-122, and the D-ITP consortium, a program of NWO that is funded by the Dutch Ministry of Education, Culture and Science (OCW).

\bibliography{hilbert_series}

\appendix
%%%%%%%%%%%%%%%%%%%%%%%%%%%%%%%%%%%%%%%%%%%%%%%%%%%%%%%%%%%%%%%%%%%%%%%%%%%%%%%%
\section{Characters}
\label{app:characters}
%%%%%%%%%%%%%%%%%%%%%%%%%%%%%%%%%%%%%%%%%%%%%%%%%%%%%%%%%%%%%%%%%%%%%%%%%%%%%%%%
In group theory, the character of a representation is equal to the trace of the matrix constituting this representation. Characters of conjugate group elements are equal, and so for simple and semi-simple Lie groups the characters can be given in terms of the Cartan matrices' parameters. \tab{characters} contains explicit expressions for the characters of all representations that we have implemented. In the fourth column we list the explicit \FORM procedures (for the Lorentz group representation) and the input for these procedures (gauge group representation) that can be used in \ECO. Note that \texttt{Q} needs to be an integer.

\begin{table*}
	\begin{tabular}{l|c|c|c|c}
	         \hline \hline
		Group & Representation & Character & \FORM procedure/input & Haar measure \\
		\hline \hline
		Lorentz & $(0,0)$ & 1 & \texttt{addScalar()}& \multirow{12}{*}{\shortstack{$\frac{1}{(2\pi\img)^2}\oint_{|y_1|=1} \frac{\df y_1}{y_1}(1-y_1^2)$ \\ \qquad $\times \oint_{|y_2|=1} \frac{\df y_2}{y_2}(1-y_2^2)$}}\\
		&$(\frac{1}{2},0)$ & $y_1 + \frac{1}{y_1}$ & \texttt{addLHFermion()}& \\
		&$(0,\frac{1}{2})$ & $y_2 + \frac{1}{y_2}$ & \texttt{addRHFermion()}& \\
		&$(\frac{1}{2},0) \oplus (0,\frac{1}{2})$ & $y_1+\frac{1}{y_1} + y_2+\frac{1}{y_2}$& \texttt{addDiracFermion}& \\
		&$(\frac{1}{2},\frac{1}{2})$ & $(y_1 + \frac{1}{y_1})(y_2+\frac{1}{y_2})$& - & \\
		&$(1,0) \oplus (0,1)$ & $y_1^2+1+\frac{1}{y_1^2} + (y_1 \leftrightarrow y_2)$ & \texttt{addFieldStrength()}& \\
		&$(2,0) \oplus (0,2)$ & $y_1^4+y_1^2+1+\frac{1}{y_1^2}+\frac{1}{y_1^4} + (y_1 \leftrightarrow y_2)$ & \texttt{addGravity()}& \\
		&$(\frac{3}{2},\frac{1}{2})$ & $(y_1^3+y_1+\frac{1}{y_1}+\frac{1}{y_1^3})(y_2+\frac{1}{y_2})$ & - & \\
		&$(\frac{1}{2},\frac{3}{2})$ & $(y_1+\frac{1}{y_1})(y_2^3+y_1+\frac{1}{y_2}+\frac{1}{y_2^3})$ & - & \\
		\hline
		$U(1)$ & charge $Q$ & $x^Q$ & \texttt{Q}& $\frac{1}{2\pi\img}\oint_{|x|=1} \frac{\df x}{x}$ \\
		\hline
		$SU(2)$ & singlet & 1 & \texttt{1}& \multirow{3}{*}{$\frac{1}{2\pi\img}\oint_{|y|=1} \frac{\df y}{y}(1-y^2)$} \\
		& fundamental/doublet & $y+\frac{1}{y}$ & \texttt{2}& \\
		& triplet/adjoint & $y^2+1+\frac{1}{y^2}$ & \texttt{3}& \\
		\hline
		$SU(3)$ & singlet & 1 & \texttt{1}& \multirow{5}{*}{\shortstack{$\frac{1}{(2\pi\img)^2}\oint_{|z_1|=1}\frac{\df z_2}{z_2}\oint_{|z_2|=1} \frac{\df z_2}{z_2}$ \\ \qquad $\times(1-z_1z_2)(1-\frac{z_1^2}{z_2})(1-\frac{z_2^2}{z_1})$}}\\
		&  fundamental/3 & $z_1+\frac{z_2}{z_1}+\frac{1}{z_2}$ &\texttt{3} & \\
		&  antifundamental/$\bar{3}$ & $z_2+\frac{z_1}{z_2}+\frac{1}{z_1}$ &\texttt{3B} & \\
		&  adjoint & $z_1z_2+\frac{z_2^2}{z_1}+\frac{z_1^2}{z_2}+2+\frac{z_2}{z_1^2}+\frac{z_1}{z_2^2}+\frac{1}{z_1z_2}$ &\texttt{8} & \\
		\hline \hline
	\end{tabular}
	\caption{The symmetry groups and the representations that we implemented. The explicit form of the characters is given together with the Haar measure in terms of the Cartan variables. In the fourth column the name of the \FORM procedure or the input for these procedures is given. Note that the input for $U(1)$ can be multiple numbers for the charges of additional $U(1)$ symmetries (when no charge corresponding to an additional $U(1)$ is provided, we assume that it is zero).}
	\label{tab:characters}
\end{table*}
 
\end{document}